\documentclass[aps,prb,twocolumn,showpacs]{revtex4-1}

\usepackage{graphicx}
\DeclareGraphicsExtensions{.png}
\DeclareGraphicsExtensions{.jpeg}

\usepackage{xcolor}
\usepackage{amsmath}
\usepackage{amssymb}

\usepackage{dcolumn}
\usepackage{bm}

\usepackage[latin1]{inputenc}
\usepackage[american]{babel}
\usepackage{latexsym}
\usepackage{float}

\begin{document}

\title{Quantum Spin Hall Effect in Twisted Bilayer Graphene}

\index{F. Finocchiaro}
\index{F. Guinea}
\index{P. San-Jose}

\author{F. Finocchiaro$^{1}$, F. Guinea$^{1,3}$ and P. San-Jose$^{2}$}
\affiliation{${}^1$IMDEA Nanociencia, Calle de Faraday 9, 28049 Madrid, Spain}
\affiliation{${}^2$ICMM-CSIC, Sor Juana Ines de La Cruz 3, 28049 Madrid, Spain}
\affiliation{${}^3$Department of Physics and Astronomy, University of Manchester, Manchester M13 9PL, United Kingdom}

\date{\today}

\begin{abstract}
Motivated by a recent experiment (Sanchez-Yamagishi \emph{et.al}, arXiv:1602.06815) reporting evidence of helical spin-polarized edge states in layer-biased twisted bilayer graphene under a magnetic flux, we study the possibility of stabilising a Quantum Spin Hall (QSH) phase in such a system, without Zeeman or spin-orbit couplings, and with a QSH gap induced instead by electronic interactions.
We analyse how magnetic flux, electric field, interlayer rotation angle, and interactions (treated at a mean field level) combine to produce a pseudo-QSH with broken time-reversal symmetry, and spin-polarized helical edge states. The effect is a consequence of a robust interaction-induced ferrimagnetic ordering of the Quantum Hall ground state under an interlayer bias, provided the two rotated layers are effectively decoupled at low energies. We discuss in detail the electronic structure, and the constraints on system parameters, such as the angle, interactions and magnetic flux, required to reach the pseudo-QSH phase. We find, in particular, that purely local electronic interactions are not sufficient to account for the experimental observations, which demand at least nearest-neighbour interactions to be included.
\end{abstract}

\maketitle


\section{Introduction}

The search for topological electronic phases and the efforts towards their experimental realisation is to date at the center of the debate in the condensed matter community, in part because of their outstanding fundamental physical manifestations, such as perfectly quantised conductivity and robustness against disorder, but also due to their potential application for quantum technologies and quantum computation \cite{Xiao:RMP10,Zhang:N05,Novoselov:N05,Halperin:PRB82,Laughlin:PRB81}. One of the most celebrated topological phases is the Quantum Spin Hall (QSH) state, that characterises time-reversal-symmetric (TRS) topological insulators in two dimensions (2D), such as HgTe quantum wells \cite{Konig:S07} or Bismuth-based compounds \cite{Sabater:PRL13,Drozdov:N14,Luo:NL15,Li:SR16}. The QSH phase consists of a 2D bulk with an inverted (topological) gap, which gives rise to topologically protected counterpropagating helical edge states related by TRS. These states remain gapless and topologically protected against spin-independent disorder as long as TRS is preserved, and are the basis of an important class of proposed quantum technologies, such as Majorana-based \cite{Fu:PRL08} quantum computation \cite{Nayak:RMP08}. 

A robust QSH phase requires, crucially, that bulk states have a sufficiently large gap. This has proved to be problematic experimentally, as most samples suffer from substantial leakage of edge states into the bulk due to disorder and imperfections, which quickly destroys their topological protection. Hence, considerable efforts are being devoted to finding the ideal material or platform to implement robust QSH phases with large bulk gaps. In most current implementations, however, the crucial parameter that controls the gap is the spin-orbit coupling of the material, which is usually not very large. Such is the case e.g. of the original Kane-Mele proposal for graphene \cite{Kane:PRL05} and most other experimental systems, including inverted quantum wells. 

A promising class of alternative QSH implementations is recently being considered, in which the topological gap is controlled by a different scale in the problem. Such is the case of graphene monolayers under strong in-plane and out of plane magnetic fields. It was predicted theoretically \cite{Abanin:PRL06,Fertig:PRL06,Lado:PRB14} and demonstrated experimentally \cite{Young:N14}, that under such conditions, graphene's unique Quantum Hall (QH) phase may be manipulated into a \emph{pseudo}-QSH phase at charge neutrality by Zeeman-polarizing the ground state, no spin-orbit coupling required. The `pseudo' qualifier refers to the absence of TRS in this scenario (unitary symmetry class A, like the QH phase, instead of the symplectic class AII of the QSH phase). The Zeeman field polarises graphene's special zero Landau level (ZLL), which leads to the coexistence of an electronic-like and a hole-like QH phases, with counter-propagating edge states. These are \emph{not} related by TRS, but nevertheless remain gapless due to their opposite spin polarisation. Note that this caveat is not a fundamental shortcoming for many applications. As an example, it has been shown that Majorana bound states may emerge in superconducting systems with broken TRS (D-class) \cite{Oreg:PRL10,Lutchyn:PRL10}, or from pseudo-QSH phases in graphene \cite{San-Jose:PRX15}. Other noteworthy systems have been recently shown to host pseudo-QSH phases with robust edge transport \cite{Ma:NC15}.

While fundamentally interesting, the above pseudo-QSH phase in monolayer graphene suffers from shortcomings of its own. Most prominent is the need of strong in-plane magnetic fields exceeding 20 Tesla, a problematic requirement when other ingredients such as superconductivity are involved in a given application (e.g. Majoranas). These large fields are necessary since, much like spin-orbit coupling in more conventional QSH implementations, the Zeeman coupling is relatively small in this system.

In this work we study a related pseudo-QSH implementation, based on (effectively) decoupled graphene bilayers in the QH regime. The proposal relies on electronic interactions to break the degeneracies of the ZLL, a mechanism demonstrated in a number of experiments \cite{Zhang:PRL06,Jiang:PRL07,Song:N10,Young:N14} that allows, by independently tuning the filling factor of each layer to opposite values $\nu=\pm 1$, to engineer a pseudo-QSH phase without the need of either spin-orbit or Zeeman couplings, and with a bulk gap controlled by electronic interactions instead. Very recently, an experimental effort has realised this pseudo-QSH implementation, albeit using a twisted graphene bilayer \cite{Sanchez-Yamagishi:16} instead of two decoupled monolayers. We demonstrate that, despite the electrical contact between the two layers in this experiment, the interlayer rotation is indeed sufficient to produce an effective layer decoupling at small magnetic fields and large rotation angles. By treating interactions within a self-consistent mean field approach, we furthermore characterise the Landau level splitting and magnetotransport properties across the full phase diagram of filling factors, paying particular attention to the important case of the pseudo-QSH phase, the associated magnetic ordering, and, finally, the role of long-range interactions, which prove to be crucial to reproduce the experimental results. We finally derive a phenomenological model for all relevant scales, which allows us to set quantitative bounds for the rotation angle, magnetic field and interaction strength to achieve the optimal pseudo-QSH regime.

This paper is organised as follows. In Sec. \ref{sec:model} we construct a physical model with the various ingredients relevant to the problem. In Sec. \ref{sec:broadening} we analyse the role of interlayer coupling on the Landau level broadening, and the connection to the Hofstadter butterfly spectrum. In Sec. \ref{sec:interactions} we explore the effect of local interactions on the ZLL, and explore the different symmetry-broken ground states that emerge at a mean field level. In Sec. \ref{sec:pseudoQSH} we combine these results with an interlayer bias, and analyse the conditions that give rise to a pseudo-QSH phase, and the corresponding phase diagram. In Sec. \ref{sec:nonlocal} we include non-local interactions, and characterize their effect on the pseudo-QSH gap. Finally, in Sec. \ref{sec:discussion} we discuss the practical implications of the results.

\section{Modelling a pseudo-QSH phase in twisted graphene bilayers}
\label{sec:model}

\begin{figure*}[t]
\centering
\includegraphics[scale=0.51, trim=5 5 5 6,clip]{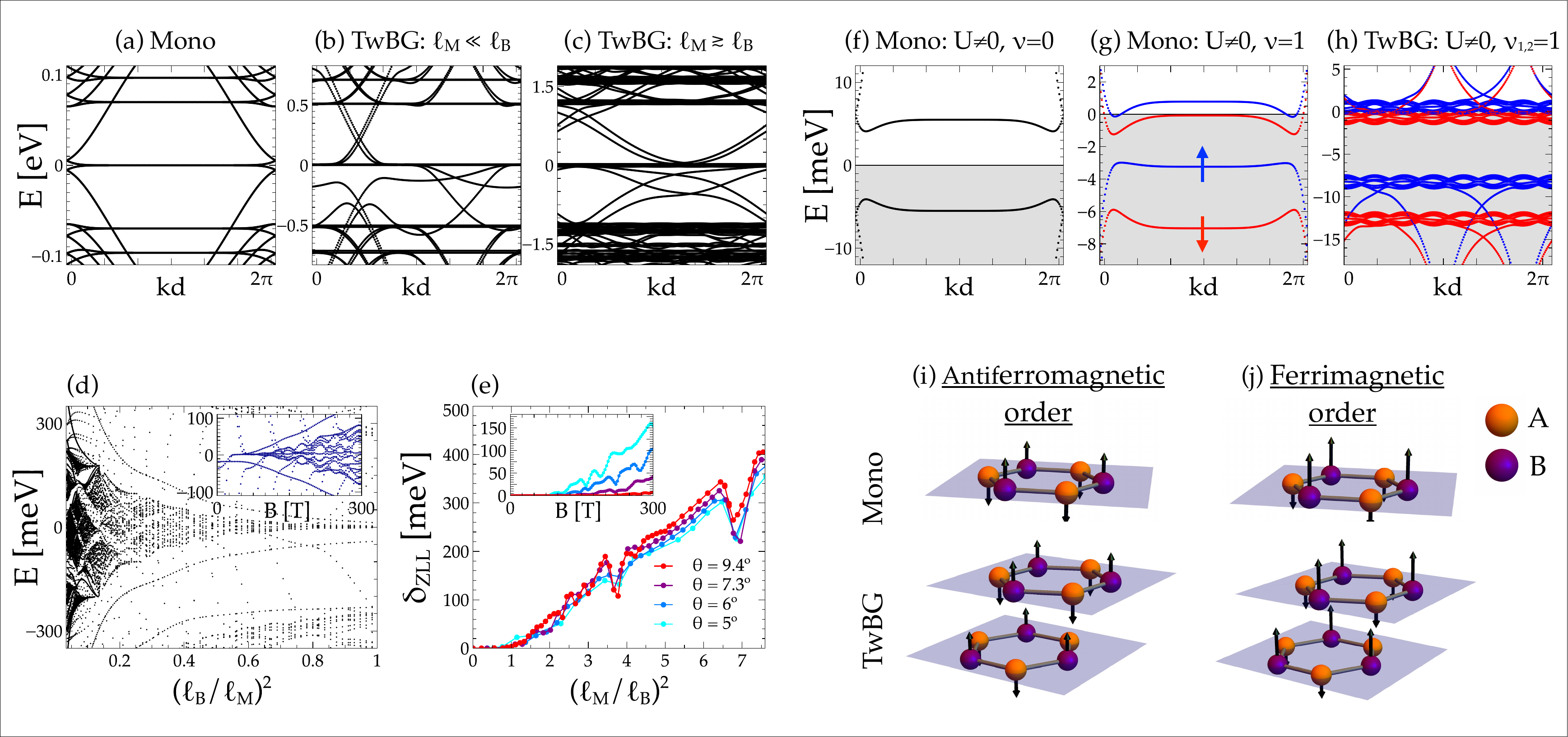}
\caption{(a) Bandstructure of a graphene monolayer nanoribbon with armchair edges in the QH regime. (b) and (c), QH bandstructure for a twisted bilayer graphene nanoribbon in the regimes $\ell_M \ll \ell_B$ and $\ell_M > \ell_B$. (d) Corresponding spectrum at fixed momentum $k = 0$ as a function of $(\ell_B/\ell_M^2$, exhibiting the fractal Hofstadter butterfly structure. Inset: Same spectrum as a function of $B$. (e) Broadening of the zero Landau level (ZLL) vs. $(\ell_M/\ell_B)^2$ for different twist angles $\theta$. Inset: Broadening as a function of $B$. (f-h) Effect of interactions on the ZLL for a monolayer and twisted bilayer. Red and blue colors indicate opposite spin polarizations. (i,j) Magnetic structure of the ground state at zero filling (antiferromagnetic) and filling $\nu=\pm 1$ (ferrimagnetic), respectively.}
\label{fig:bands}
\end{figure*}

We here develop, in steps, a model that includes all the ingredients involved in the implementation of the pseudo-QSH phase in twisted bilayer graphene (TwBG). We consider an infinite TwBG nanoribbon of width $W$, oriented along the $\hat x$ direction, and under a perpendicular magnetic field $\bm B=B\hat z$. The magnetic field and width should be large enough to drive the TwBG nanoribbon into the QH regime, i.e. $W\gg\ell_B=\sqrt{\hbar/eB}$. If we momentarily neglect the interlayer coupling and electronic interactions, and consider a gapless Dirac spectrum with velocity $v_F$, the nanoribbon under $B$ develops the well known Landau level spectrum characteristic of graphene Hall bars \cite{Neto:RMP09}, with non-dispersive Landau levels at energies $E_N=\pm \hbar v_F/\ell_B\sqrt{2n}$ ($n=0,1,2,\dots$) on each of the two decoupled layers, see Fig. \ref{fig:bands}a.
Of special interest is the $n=0$ level at zero energy, or ZLL. It has a four-fold degeneracy per layer, and spawns one electron-like and one hole-like dispersive edge states per spin at each edge and on each layer of the nanoribbon. This spectral structure for the case of decoupled monolayers is essentially independent of the specific details of the model, the type of edges and the interlayer rotation. It has been computed here assuming an armchair nanoribbon for each layer, modelled within a standard nearest-neighbour tight-binding description between $\pi$ orbitals in a honeycomb lattice \cite{Neto:RMP09}. 

\subsection{Landau level broadening in a twisted graphene bilayer nanoribbon}
\label{sec:broadening}

In actual TwBG nanoribbons, the two layers are in electrical contact, with a modulation of the interlayer coupling that follows the moir\'e pattern created by the interlayer rotation angle $\theta$ \cite{Santos:PRL07,Santos:PRB12,Bistritzer:P11}. This moir\'e pattern has a period $\ell_M=a_0/\sqrt{2(1-\cos\theta)}\approx a_0/\theta+\mathcal{O}(\theta)$, where $a_0=0.246$ nm is graphene's lattice constant and $\theta<30^\circ$. The local interlayer stacking changes continuously across the moir\'e pattern. To model the corresponding interlayer coupling, we define a tight-binding approximation beyond nearest-neighbours, with a position-dependent hopping amplitude between any two sites (within the same or different layers), 
\[
H_0=\sum_{ij} t(\textbf{r}_i-\textbf{r}_j) c_i^{\dagger}c_j.
\]
The two layers, separated by a distance $d=2.36 a_0$, are honeycomb lattices with a relative rotation $\theta$, and in a nanoribbon geometry $-W/2<y<W/2$.
The hopping function $t(\textbf{r})$ is derived within a Slater-Koster approach from the overlap of two $\pi$ orbitals, separated by a vector $\bm r=(x,y,z)$,
\begin{equation}\label{TBmodel} 
t(\textbf{r})=
\left\{ 
\begin{array}{ccc}
-\frac{x^2+y^2}{r^2}t e^{-\beta\frac{r - a}{a}}+ \frac{z^2}{r^2}t_{\perp}e^{-\beta\frac{r-d}{a}} & \mbox{if }r\leq R\\
0 & \mbox{if }r>R
\end{array}
\right.
\end{equation}
Here $t=2.7\,\mbox{eV}$ and $t_{\perp}=0.178t$ are the intra- and inter-layer hoppings respectively, $a=a_0/\sqrt{3}$ is the Carbon-Carbon distance, and $r=|\bm r|$ \cite{Moon:PRB13}.
The dimensionless parameter $\beta\approx 3.14$ controls the range of the hopping amplitude. $R$ is a cutoff distance for the hoppings, introduced for numerical purposes, and chosen large enough ($R = 4 a_0$) so that the spectrum is independent of $R$. The magnetic flux is again incorporated into the hopping through a Peierls phase, $t(\bm r_i-\bm r_j)\rightarrow \exp\left[i\bm A(\bm r)\cdot(\bm r_i-\bm r_j)\right]t(\bm r_i-\bm r_j)$, with the vector potential chosen in the Landau gauge $\bm A(\bm r)=-B\hat {\bm y}$, and evaluated at the hopping midpoint $\bm r=(\bm r_i+\bm r_j)/2$.

The resulting non-interacting model Hamiltonian $H_0$ is translationally invariant, so one may once more compute its bandstructure as a function of Bloch momentum $k d$ along the nanoribbon, where $d$ is the length of the nanoribbon unit cell, \footnote{This requires the interlayer rotation to yield commensurate structures, i.e. a finite $d$, so that $\theta$ is constrained to a discrete set of values. It has been shown, however, that with a smooth hopping model as the one used here, the resulting electronic structure have exponentially small commensurate-incommensurate effects \cite{Santos:PRB12}.} and compare it to the case of a decoupled bilayer. At zero magnetic flux, it has been shown that the interlayer coupling in TwBG has a negligible effect at low energies, as long as the rotation angle is large, $\theta\gtrsim 1^\circ$ \cite{Bistritzer:P11}.
This is a result of the momentum mismatch between the Dirac cones in each layer, which are shifted apart by the $\theta$ rotation. For small but finite magnetic field $B$, such that $\ell_M\ll \ell_B\ll W$, this is still the case, and the two layers remain effectively decoupled at low energies. Hence, Landau levels close to zero energy strongly resemble those of the decoupled bilayer, and in particular remain perfectly flat, see Fig. \ref{fig:bands}b. The dispersion of edge states depends on the termination of each of the layers, which is different from armchair due to the $\theta$ rotation. Otherwise, the bandstructure is a two-fold degenerate version of graphene's QH spectrum. As the magnetic field is increased further, the conventional picture of decoupled layers breaks down, and Landau levels acquire a \emph{finite width} $\delta_{ZLL}$, see Fig. \ref{fig:bands}c. This happens as soon as the magnetic length becomes smaller than the moir\'e period, $\ell_M\gtrsim \ell_B$, i.e. when the magnetic flux per moir\'e supercell becomes comparable to the flux quantum. This marks the onset of the Hofstadter butterfly regime predicted in moir\'e superlattices \cite{Bistritzer:PRB11}, which severely disrupts the conventional Landau level spectrum. The emergence of the Hofstadter butterfly in the nanoribbon as $B$ is increased at a fixed angle, that is as $\ell_B/\ell_M$ is decreased, is demonstrated in Fig. \ref{fig:bands}d, with the corresponding ZLL broadening $\delta_{ZLL}$ for different angles shown in Fig. \ref{fig:bands}e.

\begin{figure*}
\centering
\includegraphics[scale=0.48, trim=7 6 6 6,clip]{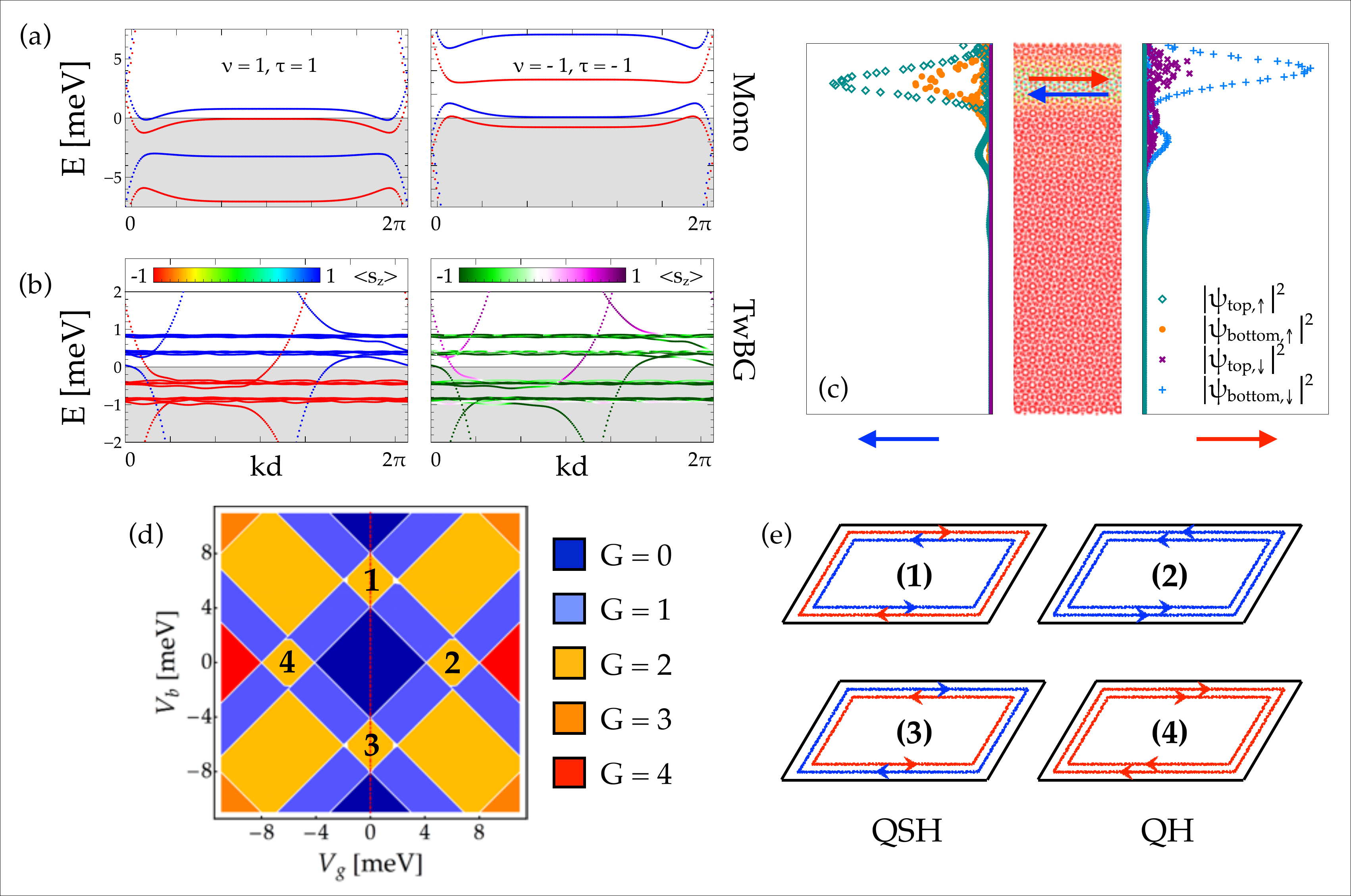}
\caption{(a) Pseudo-QSH bandstructure of two decoupled graphene monolayer with armchair edges, at fillings $\nu = 1$ and $\nu=-1$. $\tau = \pm 1$ stands for the top and bottom layer respectively.  (b) Pseudo-QSH bandstructure for a TwBG nanoribbon at the same filling, induced by a bias $V_b$ between the two layers. Colors in the right (left) panel indicate spin (layer) polarization. Note that the ribbon has two edges, so that there are two states per edge. (c) Spin- and layer-resolved spatial density of the edge states along a TwBG semi-infinite plane in the pseudo-QSH regime. (d) Phase diagram of the two-terminal conductance $G$ in units of $e^2/h$ for fixed values of $B= 4$ T, $U=1.5 t$ and $U_2=0.28 U$, as a function of the total gating $V_g$ and interlayer bias $V_b$. (e) Sketch of the edge states configuration in the four FI-ordered phases marked as (1-4) in (d), with red and blue denoting opposite spin polarizations.}
\label{fig:phasediagram}
\end{figure*}

\subsection{Interactions in the Quantum Hall regime}
\label{sec:interactions}

So far we have discussed the non-interacting QH bandstructure. We now discuss the effect of electron-electron interactions. We first consider completely screened interactions, in the form of a purely on-site Hubbard term. The Hamiltonian then becomes $H=H_0 + H_{int}$ where $H_{int}=U\sum_{i}n_{i\uparrow}n_{i\downarrow}$. To compute the effects of $H_{int}$ we employ a standard mean field decoupling, so that $H_{int}$ is approximated by a self-consistent $H_{int}^{MF}=U\sum_{i}\left[n_{i\uparrow}\left\langle n_{i\downarrow}\right\rangle + \left\langle n_{i\uparrow}\right\rangle n_{i\downarrow} \right]+E_U$, with $E_U$ an unimportant constant \footnote{Note that Fock terms are not required in this decoupling, as $H_0$ is SU(2) symmetric. The Hartree decoupling then merely amounts to a specific choice of magnetization direction for symmetry breaking.}.

In monolayer graphene, at half filling and for high enough perpendicular magnetic fields, interactions are able to break the $SU(4) = SU(2)\times SU(2)$ spin and sublattice symmetry of the ZLL \cite{Kharitonov2:PRB12,Kharitonov:PRB12,Kotov:RMP12,Abanin:PRL06,Nomura:PRL06,Gusynin:PRB06,Alicea:PRB06,Herbut:PRB07,Fuchs:PRL07,Sheng:PRL07,Jung:PRB09,Shylau:PRB11}. Within the above Hubbard model, the symmetry breaking happens already at the self-consistent mean field level, which at zero filling $\nu=0$ predicts a sublattice-splitted $SU(2)$-symmetric ZLL with a gap $\Delta_{AF}$ and characterized by an antiferromagnetically (AF) ordered ground state \cite{Kharitonov2:PRB12,Kharitonov:PRB12,Abanin:PRL06,Lado:PRB14,Herbut:PRB07}, consistent with experiments \cite{Young:N14}. Figure \ref{fig:bands}f shows the $\nu=0$ Hubbard spectrum for the monolayer at small energies, with the AF ground state polarization depicted in Fig. \ref{fig:bands}i.  This AF ground state, with opposite spin polarization on different A/B sublattices, hosts no protected edge states, and is thus fully insulating. As the filling factor is increased to $\nu=1$, the remaining $SU(2)$ symmetry of the split ZLL level is  broken, see Fig. \ref{fig:bands}g. A small $\nu=1$ gap $\Delta_{FI}$ opens, and the corresponding bulk ground state becomes ferrimagnetic (FI), with uncompensated spin polarization between A/B sublattices, see Fig. \ref{fig:bands}j. Protected edge states emerge within the FI gap, which, remarkably, are perfectly spin-polarized (red and blue in Fig. \ref{fig:bands}g denote opposite spin polarization $\langle s_z\rangle=\pm 1$). The spin polarization of these edge states stems entirely from interactions, as no Zeeman coupling is included here.

As in the non-interacting case, the spectrum of the TwBG nanoribbon exhibits similar phenomenology as the two decoupled layers. This is true as long as $\ell_B>\ell_M$. Otherwise the Landau level broadening $\delta_{ZLL}$ may exceed the relevant gap, be it $\Delta_{AF}$ or $\Delta_{FI}$, and the bulk of the system then becomes metallic, as shown in \ref{fig:bands}h.

\subsection{Pseudo-QSH under an interlayer bias}
\label{sec:pseudoQSH}

We emphasise once more, in relation to the possibility of a pseudo-QSH phase (with helical counterpropagating edge states of opposite spins), the fact that within the $\nu=1$ gapped phase, each layer hosts one \emph{spin-polarized} state at a given edge, \footnote{The precise statement is that the total number of right-moving minus left-moving states per edge and layer is one} as sketched in panel (2) of Fig. \ref{fig:phasediagram}e, without the need of a Zeeman field. This is one of the key ingredients that allow the bilayer to be coaxed into a pseudo-QSH phase. A second ingredient is that the spin orientation and edge state propagation becomes inverted for negative filling $\nu=-1$, see panel (4) of Fig. \ref{fig:phasediagram}e. Thus, a decoupled bilayer tuned to $\nu=1$ in one layer and $\nu=-1$ in the other results in an implementation of a pseudo-QSH phase.

Independent filling of two decoupled graphene monolayers, separated by an insulator such as hexagonal Boron Nitride (hBN), may in practice be achieved using electrostatic gating with a top and bottom gate. Together, they can be used to induce independent shifts $V_1$ and $V_2$ in the Fermi energy of bottom (1) and top (2) layers. Instead of treating the associated electrostatic and screening problem as a function of gate potentials, we simply express the associated energy shifts $V_{1,2}$ in terms of an actual interlayer bias $V_b=V_2-V_1$, and overall gating $V_g=(V_1+V_2)/2$. These enter $H_0$ as a term of the form $\sum_i c^\dagger_i(V_b\tau_z+V_g\tau_0)c_i$, where $\tau_i$ are Pauli matrices acting in layer space. By adjusting $V_b$ and $V_g$ we can control the filling factor of each layer independently. At $\nu_2=-\nu_1=1$, each layer of the decoupled bilayer develops the spectrum depicted in Fig. \ref{fig:phasediagram}a, which, taken together, indeed corresponds to a pseudo-QSH phase, with counter propagating edge states of opposite spins as sketched in panel (1) of Fig. \ref{fig:phasediagram}e. Naturally, the decoupling between the layers makes these edge states gapless, as interlayer scattering is not allowed, either through hopping or electronic interactions. A non-trivial question is whether these states also emerge at low energies in a coupled TwBG nanoribbon with generic edges and interactions, and whether they remain gapless like in a proper QSH, despite the lack of time reversal symmetry, as experimentally suggested \cite{Sanchez-Yamagishi:16}.

We compute the spectrum of a coupled TwBG nanoribbon at the special $\nu_2=-\nu_1=1$ configuration above, again using only on-site Hubbard interactions. In general, the interlayer coupling precludes a definition of the filling factor of each layer independently, as Landau levels may now have a finite participation on both. The effective layer decoupling at low energies due to the interlayer rotation, however, makes it possible to reach a pseudo-QSH regime equivalent to Fig. \ref{fig:phasediagram}a, provided $\ell_B\gg\ell_M$. The typical spectrum is shown in Fig. \ref{fig:phasediagram}b, where we see that, as long as the residual Landau level broadening is smaller than the FI gap $\Delta_{FI}$, a robust pseudo-QSH phase is stabilised. Counterpropagating edge states remain perfectly spin-polarized ($\langle s_z\rangle=\pm 1$ shown as red and blue in the left panel), which guarantees that they remain gapless, provided all perturbations, imperfections or disorder at the edge are spin-independent (a requirement also in conventional QSH phases). In contrast, the layer polarization of counterporpagating edge states, denoted as $\langle\tau_z\rangle$, is quite strong, but not perfect. This is depicted on the right panel of Fig. \ref{fig:phasediagram}b, and is confirmed by the edge states wavefuncions plotted in Fig. \ref{fig:phasediagram}c.

The full phase diagram for arbitrary layer filling is computed next. As a function of $V_g$, which controls $(\nu_1+\nu_2)/2$, and $V_b$, which controls $\nu_2-\nu_1$, we find the phases shown in Fig. \ref{fig:phasediagram}d. \footnote{The phase diagram is computed in the limit of decoupled layers for simplicity, but we have checked that interlayer coupling merely broadens phase boundaries. The interaction also includes a nearest-neighbour terms $U_2=0.28U$, which changes the relative sizes of each region as discussed below.} The color code represents the two-terminal conductance $G$ (in units of $e^2/h$) of each phase, assuming purely spin-independent scattering. $G$ is is quantized as transport always occurs through edge states. Around the origin, with $G=0$, we have the AF phase. The regions (1-4) correspond to FI ground states, with $|\nu_1|=|\nu_2|=1$ and a conductance $G=2\,e^2/h$ arising from the edge states sketched in Fig. \ref{fig:phasediagram}e. Phases (1) and (3) are pseudo-QSH, while (2) and (4) are spin-polarized QH phases. A similar phase diagram has been experimentally measured in Ref. \onlinecite{Sanchez-Yamagishi:16}. Deviations occur mostly in the shape of each region. This is expected, as the control parameters in the experiment are not the actual potentials $V_b$ and $V_g$ on the two layers, but rather actual applied potentials before interlayer screening.
We mention also that in Ref. \onlinecite{Sanchez-Yamagishi:16} the two FI regions corresponding to opposite filling factors ((1) and (3) in  Fig. \ref{fig:phasediagram}d) display conductances ranging between 0.8 and 1.3 $e^2/h$, that is around half the value that is expected according to our model. The precise reasons for such deviation from the ideal $G=2\,e^2/h$ are yet unclear, experimentally, but are likely due to a residual spin-flip scattering amplitude associated to spin-dependent disorder, non-collinear edge-magnetization \citep{Lado:PRB14}, and/or gapless collective spin excitations of the ground state.

\subsection{Non-local interactions}
\label{sec:nonlocal}

\begin{figure}[t]
\centering
\includegraphics[width=\columnwidth, trim=8 6 6 9,clip]{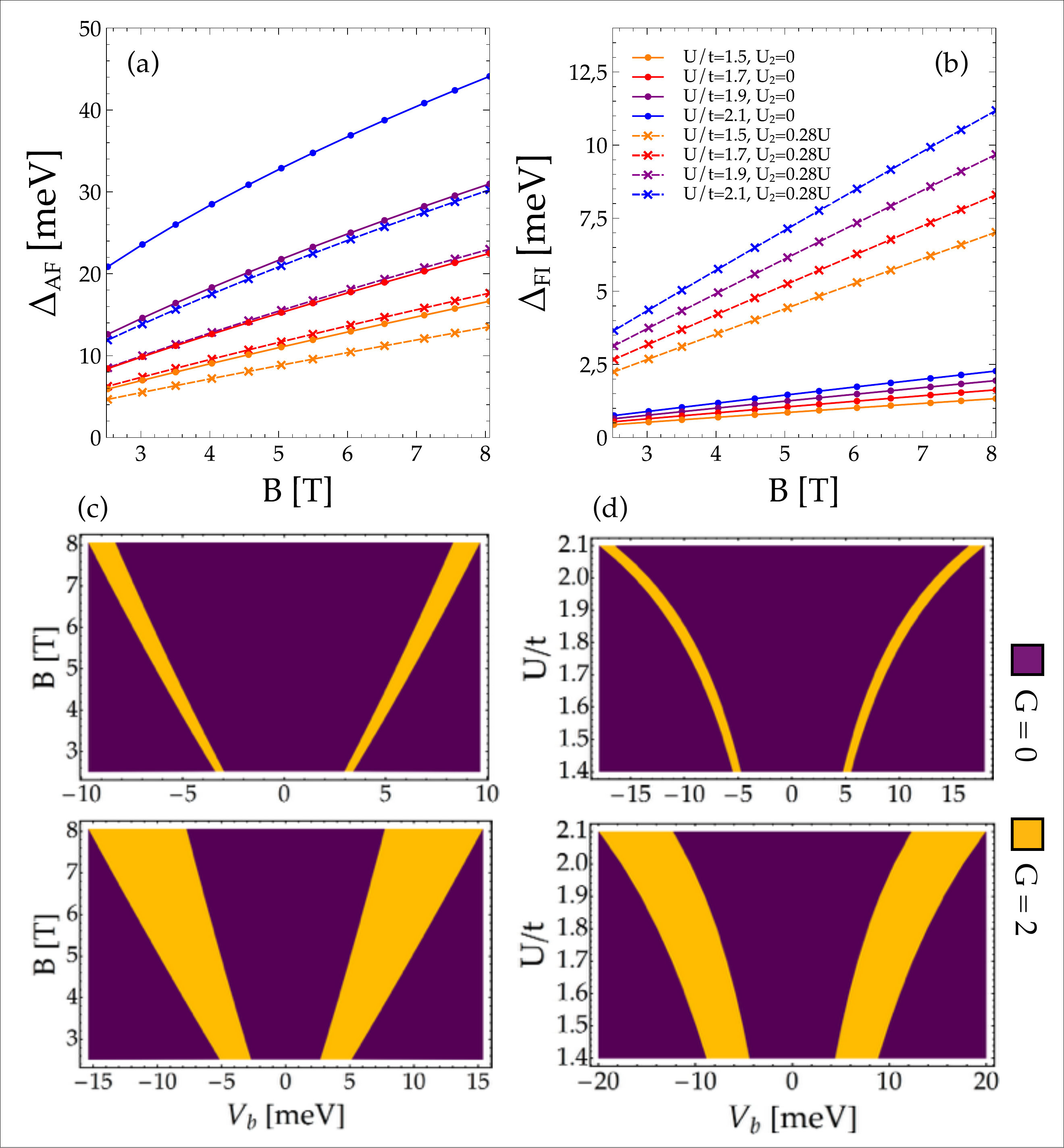}
\caption{(a) and (b) Antiferromagnetic and ferrimagnetic gaps $\Delta_{AF}$ and $\Delta_{FI}$ as functions of the magnetic field $B$ for different values of $U$. The cases with (dashed) and without (solid) nearest-neighbour interactions $U_2$ are compared. (c) and (d) Maps of the conductance $G(V_b)$ at $V_g=0$ (red vertical cut in Fig. \ref{fig:phasediagram}d) as a function of $B$ (panel (c), with $U/t=1.5$) and as a function of $U/t$ (panel (d), with $B=5$ T). Top and bottom rows correspond to $U_2=0$ and $U_2=0.28 U$, respectively. }
\label{fig:gaps}
\end{figure}

An important deviation between the preceding simulations, using purely on-site Hubbard interactions, and the experimental phase diagram is the relative size of the the FI phases (1-4) respect to the AF phase at $\nu_1=\nu_2=0$. These phases are comparable in size in the measured phase diagram, which implies that $\Delta_{FI}$ and $\Delta_{AF}$ are themselves comparable. We cannot reproduce this result using purely on-site interactions, which yield $\Delta_{FI}\ll \Delta_{AF}$. In fact, the computation of the phase diagram in Fig. \ref{fig:phasediagram}d, with AF/FI proportions closely matching the experiment, requires the inclusion of a nearest-neighbour interactions $U_2$ 
\begin{equation}
H_{int}'=U\sum_i n_{i\uparrow}n_{i\downarrow} +\frac{ U_2 }{2}\sum_{\stackrel{\langle ij\rangle}{\sigma\sigma'}} n_{i\sigma}n_{j\sigma'}
\end{equation}
To treat these non-local interactions, we employ once more a mean field decoupling of this interaction, taking care to include the non-local Fock term $U_2\sum_{\langle ij\rangle,\sigma\sigma'}[\langle n_{i\sigma}\rangle n_{j\sigma'}- c_{i\sigma}^{\dagger}c_{j\sigma}\langle c_{j\sigma}^{\dagger} c_{i\sigma}\rangle]$.

A finite $U_2$ strongly enhances $\Delta_{FI}$. 
In Fig. \ref{fig:gaps}(a,b) we show the $\Delta_{AF}$ and $\Delta_{FI}$ gaps using the two interaction models (solid and dashed curves for $H_{int}$ and $H_{int}'$, respectively).
We find that finite values of $U_2$ of the order of $0.2U-0.3 U$ significantly increase the $\Delta_{FI}$ gaps, while slightly reducing the AF gaps $\Delta_{AF}$. In panels (c) and (d) we likewise compare a $V_g=0$ cut of the phase diagram as $B$ and $U$ are varied, using a $U_2=0$ model (top row) versus a $U_2 \neq 0$ model (bottom row). This type of cuts were measured in Ref. \onlinecite{Sanchez-Yamagishi:16} as a function of $B$.
We find good quantitative agreement with the experiment using $U_2=0.28 U$, and conversely can rule out the $U_2=0$ model. We thus conclude that while purely local interactions in graphene may account for the type of symmetry breaking of the QH regime observed experimentally, a quantitative agreement requires sizeable non-local interactions, at least up to nearest-neighbours. This observation matches ab-initio calculations that predict non-negligible interactions beyond the local Hubbard U model \cite{Wehling:PRL11}.

\section{Discussion}
\label{sec:discussion}

We finish by presenting analytical expressions for the curves previously derived and establishing quantitative bounds required for the onset of a robust pseudo-QSH regime in twisted bilayers, an important issue in view of future implementations of this regime within quantum technological applications, such as the generation of Majorana bound states in twisted graphene bilayers without spin-orbit or Zeeman couplings. In this regard, we expect that a twisted graphene bilayer might be a superior choice over monolayers separated by thin insulators such as e.g. hBN, since the former has the minimum possible thickness, and interlayer superconducting correlations required for Majoranas may thus be easier to establish.

As discussed, the essential condition that must be satisfied for the pseudo-QSH to emerge at $\nu_1=-\nu_2=1$ is that the Landau level broadening does not exceed the FI gap, $\delta_{ZLL}<\Delta_{FI}$. From the simulations of Fig. \ref{fig:bands}e, we obtain a phenomenological equation for $\delta_{ZLL}$ that reads
\begin{equation}
\delta_{ZLL}\approx \left\lbrace
\begin{array}{ll}
0 & \mbox{if } (\ell_M/\ell_B)^2 \leq x_{c} \\
\alpha \left[(\ell_M/\ell_B)^2-x_{c}\right] & \mbox{if } (\ell_M/\ell_B)^2 > x_{c} \\
\end{array}
\right.
\label{eq:deltaZLL}
\end{equation}
with $\alpha\approx 57$ meV and $x_{c}\approx 0.9$. Such a fit is valid for every twisting angle $\theta \gtrsim 0.64$ for fields up to 10 T, that is angles well within the range for which the decoupled layer approximation considered in the paper holds ($\theta \gtrsim 1.89 ^\circ$).
On the other hand $\Delta_{FI}$ (Fig. \ref{fig:gaps}b) can be accurately fitted to $\Delta_{FI}=\beta (U-\delta U) B$ for $U>\delta U$ (zero otherwise), where parameters $\beta$ and $\delta U$ depend on the choice of $U_2$. For $U_2=0$, we have $\beta = 7.2 \cdot 10^{-5}$ $\mbox{T}^{-1}$, $\delta U = 1.73$ eV, while for the value adjusted to the experiment $U_2=0.28U$, we have $\beta = 3.2 \cdot 10^{-4}$ $\mbox{T}^{-1}$ and $\delta U = 1.00$ eV.

The relevant value of $U$ is largely uncertain.
Here we choose a reasonable value of $U=1.8 t\approx 5$ eV. Then, at a field $B$ of 1 Tesla ($\ell_B=25.66$ nm), we have an FI gap $\Delta_{FI}\approx 1.2$ meV (14 K) whereas at a field of 10 Tesla ($\ell_B=8.1$ nm), the gap is of $\Delta_{FI}\approx12.3$ meV (140 K). This scale sets the maximum temperature at which QSH phenomena could in principle be visible in this system, and comes out much larger than in any other QSH platform based on spin-orbit coupling.
The condition $\delta_{ZLL}<\Delta_{FI}$ then becomes a constraint on the twist angle $\theta>\theta_c$ for the pseudo-QSH phase, since $\theta$ enters Eq. \eqref{eq:deltaZLL} through $\ell_M$. The Landau regime $\delta_{ZLL}\approx 0$ of virtually decoupled layers is reached for $(\ell_M/\ell_B)^2 \leq x_{c}$, which corresponds to $\theta > 1.83 ^\circ$ at 10 T, or $\theta > 0.58 ^\circ$ at 1 T.
This implies that for the range of fields $B<10 T$ most relevant to realistic experiments, an angle $\theta\gtrsim 2^\circ$ is already guaranteed to satisfy $\delta_{ZLL}\approx 0<\Delta_{FI}$. Note that this $\theta\gtrsim 2^\circ$ also corresponds to the high-angle regime with weakly renormalized Fermi energy in twisted graphene bilayers \cite{Santos:PRL07}.
\section{Conclusion}
To summarize, we have characterised theoretically the electronic structure of twisted bilayers in the Quantum Hall regime, including interactions and interlayer bias. We have found that, in agreement with previous results for monolayers \cite{Abanin:PRL06}, the SU(4)-symmetric zero Landau level at each layer experiences spontaneous symmetry breaking due to the interactions, with either an antiferromagnetic or ferrimagnetic ground state, depending on the filling. We demonstrate that at realistic magnetic fields, the interlayer coupling does not qualitatively modify this picture. The ferrimagnetic phase, in combination with an interlayer bias, allows for the implementation of a pseudo-QSH phase with helical edge states for conveniently tuned layer filling factors. This implementation of a QSH phase is unique, in that neither spin-orbit nor Zeeman coupling is involved, unlike in all previous approaches. Despite the broken time-reversal symmetry, the pseudo-QSH edge states remain gapless in this system for arbitrary spin-independent disorder, like in conventional QSH. Their spin-polarization is a consequence of interactions, which control the scale of the associated QSH gap.
  
Our theoretical description matches the recent measurements in this system \cite{Sanchez-Yamagishi:16}, which allow us to constrain the relevant set of microscopic parameters, such as the interaction model and the interlayer rotation angle. In particular, we find that non-local interactions beyond the Hubbard model are required to explain the experimental results. We note that the problem with fully unscreened non-local interactions is considerably more subtle, and has been predicted to give ferromagnetic ground states in graphene \emph{monolayers} at charge neutrality, with Luttinger liquid-like helical edge states \cite{Fertig:PRL06,Brey:PRB06a}. Such a pseudo-QSH phase is different from the one discussed here, and has not yet been demonstrated experimentally.

Given the fundamental importance of QSH phases in the emergent field of quantum technologies, and the substantial advantages and potential of the graphene-based implementation discussed here, we expect twisted graphene bilayers in the Quantum Hall regime to attract much interest in this field, particularly towards a novel implementation of topologically protected Majorana qubits.

\acknowledgements

We are grateful to J. Lado, F. De Juan and R. Aguado for fruitful discussions. F. F. and F. G. acknowledge the financial support by Marie-Curie-ITN Grant No. 607904-SPINOGRAPH. P.S-J. acknowledges financial support from the Spanish Ministry of Economy and Competitiveness through the Ram\'on y Cajal project RYC-2013-14645 and Grant No. FIS2015-65706-P (MINECO/FEDER).

\appendix

\bibliography{biblio}

\end{document}